# Phenomenological description of spread of Covid–19 in Italy: people mobility as main factor controlling propagation of infection cases

Corrado Spinella[a] and Antonio Massimiliano Mio[b]

(a) Dipartimento di Scienze Fisiche e Tecnologie della Materia, Consiglio Nazionale delle Ricerche, Piazzale Aldo Moro 7, 00185 Rome (Italy)
Email: corrado.spinella@cnr.it

(b) Istituto per la Microelettronica e Microsistemi, Consiglio Nazionale delle Ricerche, Ottava Strada 5, 95121 Catania (Italy)
Email: antonio.mio@cnr.it

## Abstract

The spread of the coronavirus (COVID–19), starting in late 2019, has determined in Italy several interventions aimed to prevent saturation of the health system. We have examined the effects of such measures by proposing a mean–field model describing the spread of the infection based on a simple diffusion process where all the observable variables (number of people still positive to the infection, hospitalized and fatalities cases, healed people, and total number of people that has contracted the infection) depend on average parameters, namely diffusion coefficient, infection cross–section, and population density. Although this model is less sophisticated than other models in the literature, it allows us to directly relate the trend of the epidemic statistical information (hospitalized cases, number of fatalities, number of infected people, etc.) to a well defined observable physical quantity: the average number of people that any individual meets per day. The model fits very well the epidemic data, and allows us to strictly relate the time evolution of the number of hospitalized case and fatalities of the outbreak to the change of people mobility, consequent to the implementation of progressive restrictions in Italy, running until the present days (November the 15$^{th}$, 2020).

# Introduction

The new coronavirus severe acute respiratory syndrome coronavirus 2 (SARS–CoV–2), initially started in the city of Wuhan, China,[1–4] has transformed into a pandemic that has affected a large number of countries around the world.[5–7] In Italy, as of November the 15th, 2020, a total of 1,178,529 cases of coronavirus disease 2019 (COVID–19) and 45,229 deaths have been confirmed.[8] Models are extremely useful to identify physical key parameters influencing the spread of infection and thus taking appropriate measures to limit serious consequences of the influenza/SARS pandemics.[3,4,7,9–14] In this work we propose a model based on the assumption that spreading of viral infection can be described by a simple diffusion process, by assuming that all the state variables of the analyzed system change in a continuous way.

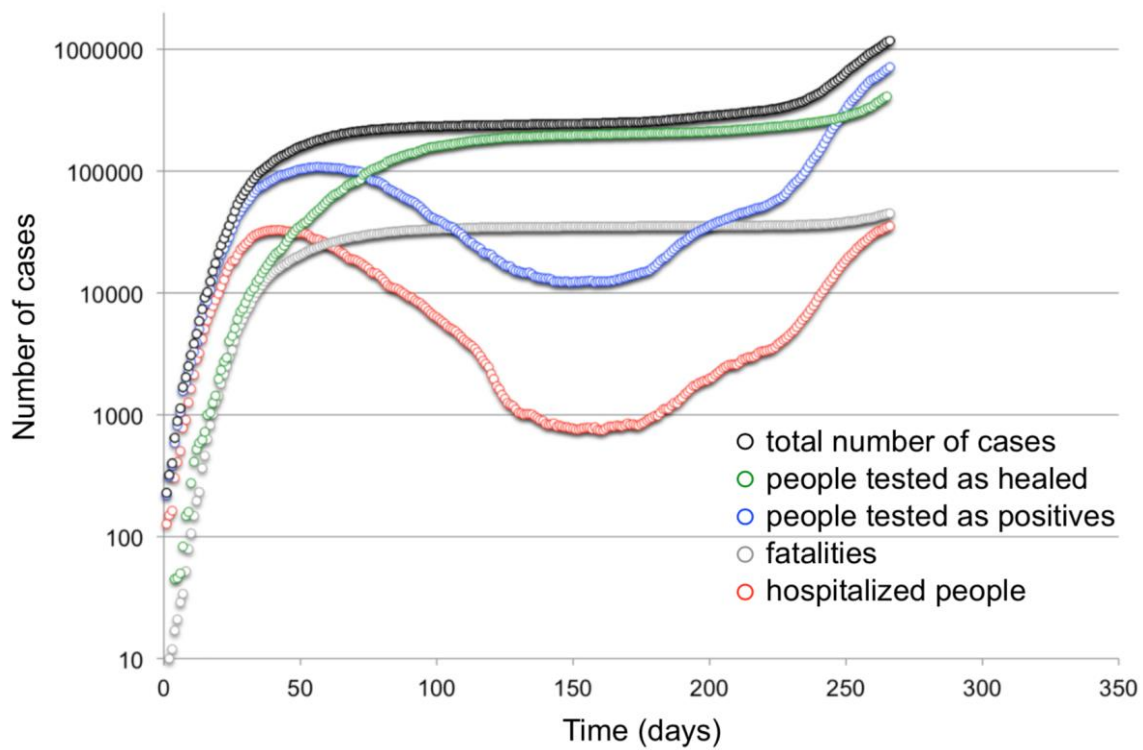

Fig. 1 Experimental data of hospitalized people (red circles), fatalities (grey circles), people tested as active positive to the viral infection (blue circles), people tested as healed (green circles), total number of cases (black circles) since the beginning of the Covid–19 outbreak in Italy (February the 24th, 2020).

Compare to other descriptions, our approach allows us to get a fast feedback of the restriction measures adopted to reduce people mobility and a prediction of the evolution scenarios on the basis of a fit to the experimental available data shown in a semi–logarithmic plot as a function of time in Fig. 1. The data refer to the numbers of the Covid–19 outbreak in Italy: people tested as active positive (blue circles), people tested as healed (green circles), hospitalized people (red circles), fatalities (grey circles), and total number of those who has been infected (black circles), since February 2020, the 24th (the day the first case of Covid–19 infection was detected in an Italian citizen). After the initial sudden increase of the number of cases, Italy implemented measures aimed to limit people mobility from March 2020 the 9th to June 2020 the 14th (day range from the 15th to the 112th in Fig. 1). The consequence of such measures was a significant slowdown of the outbreak diffusion, following by a decrease in the number of positive and hospitalized cases until day 150th. In the absence of measures strongly limiting people mobility, outbreak started to propagate again, at increasing rate, until today.

**Theoretical description**

The model we propose for describing time evolution of the total number of infected people, positive cases, healed people, deaths, and hospitalized people, during the Covid–19 outbreak is based on an mean–field approximation consisting in the assumption that the probability for an individual to contract the infection is uniformly proportional to the concentration $p$ of positive circulating cases, to a diffusion coefficient $D$, equal to the surface area covered on average by the individual in their movements in a day, and to a cross–section $\sigma^2$ measuring the probability of a single infection event ($\sigma^2 = \pi R^2$, $R$ being the average distance within which a positive individual may infect a healthy one). Under this hypothesis, at any instant $t$, the increase $dp$ of people positive for viral infection in the time interval $dt$ can be expressed as:

$$\frac{dp}{dt} = \rho D \sigma^2 (\rho - c) p - \frac{dg}{dt} - \frac{dm}{dt} \qquad (1)$$

where $\rho$ is the density of inhabitants ($\rho = 200 \text{ km}^{-2}$ for Italy), $g$ is the healed people concentration, $m$ is the concentration of fatalities, and $c$ is the total concentration of those who have contracted the virus at the time $t$. Here, we are also assuming that healed individuals reach virus immunity lasting a characteristic time much longer than the other considered time scales.

A fraction $f$ of the new positive cases requires hospital care and, consequently, the concentration $r$ of hospitalized people will change, in the time interval $dt$, by a quantity $dr$ given by:

$$\frac{dr}{dt} = f \rho D \sigma^2 (\rho - c) p - \left(\frac{q}{\tau_1} + \frac{1-q}{\tau_2}\right) r \qquad (2)$$

where we further consider that $r$ diminishes, in the same time interval $dt$, because a fraction $q$ of hospitalized people die in a characteristic time $\tau_1$, whilst the complementary fraction $(1 - q)$ heal in a characteristic time $\tau_2$. As a consequence, the concentration $m$ of fatalities varies with time according to the following equation:

$$\frac{dm}{dt} = \frac{q}{\tau_1} r \qquad (3)$$

While the fraction $f$ of positive people is hospitalized, the fraction $(1 - f)$ does not exhibit serious symptoms until complete healing. Their concentration $s$ will vary on time according to the following relationship:

$$\frac{ds}{dt} = (1-f)\rho D\sigma^2(\rho - c)p - \frac{s}{\tau_3} \qquad (4)$$

$\tau_3$ being the characteristic time toward the healing for these individuals. As reported in Ref. 15, this characteristic time is typically larger than $\tau_2$, i.e. the one used for describing time dependent healing of the most severe hospitalized cases (Eq. 2). As a consequence, the time dependence of the concentration $g$ of healed people changes in time according to:

$$\frac{dg}{dt} = \frac{s}{\tau_3} + \frac{(1-q)r}{\tau_2} \qquad (5)$$

Finally, the total concentration $c$ of those who have contracted the infection will vary on time according to the following relationship:

$$\frac{dc}{dt} = \rho D\sigma^2(\rho - c)p \qquad (6)$$

Our description is based on the assumption that the dynamics of all the observable variables, $p, r, m, g, c$, can be described in terms of the time dependence of the diffusion coefficient $D = D(t)$.

It should be noticed that the data of people tested as positive or healed (blue and green circles in Fig. 1) represent only a small fraction of the real corresponding concentration values, since they refer to the cases actually detected through the adopted testing procedure (swabs), restricted to a defined relatively small sample of the entire population. We believe that these sources of uncertainty could be partly mitigated by estimating the model parameters from a fit focused only to the data representing the time evolution of hospitalized and fatality cases (red and grey circles in Fig. 1), reported by "Dipartimento della Protezione Civile".[8]

In order to reduce the number of free parameters, we decided to fix the healing characteristic times, $\tau_2$ and $\tau_3$, to the values experimentally determined in Ref. [15] ($\tau_2 = 20$ days and $\tau_3 = 14$ days, respectively). It should be also considered that the contribution of $\sigma^2$ (the cross–section for the infection event) is not separated from the one of the diffusion coefficient $D$, since Eqs. 1, 2, 4, 6 contains only the product $\sigma^2 D$. In particular, we have set $\sigma^2 = 3.14$ m² (corresponding to $R = 1$ m), and, then, the only free parameters that remain to be determined by the fit to the experimental data are $f, q, \tau_1$.

As we have already pointed out, the model is also based on the knowledge of the time dependence of the diffusion coefficient $D$ that strongly depends on the adopted restriction measures to the people mobility. Information about the form of the function $D = D(t)$ can be inferred by admitting that $D$ can vary while calculating $r$ (Eq. 1 to 6), point by point, along the whole integration interval, in order to minimize the difference with the corresponding experimental point (we imposed that such a difference keeps lower than 1%). For the calculations we imposed the following initial conditions: $r_0 = 127$ (i.e. the experimental point at $t = 0$), $p_0 = r_0/(1-f)$, $c_0 = p_0$, $g_0 = 0, m_0 = 0$.

The obtained values of $D$, determined as a function of different sets of the $f, q, \tau_1$ parameters, are plotted in Fig. 2 as a function of time (red curves). We notice that $D$ depends only very weakly on $f, q, \tau_1$. Its behavior is characterized by a strong decrease, extending until June, the 14$^{th}$ and corresponding to the end of the initial stages of the mobility restriction measures, followed by a smooth increase lasting till October, the 1$^{st}$, and by an even more significant increase starting after October, the 1$^{st}$ and extending till October, the 25$^{th}$. After that date, $D$ starts to decrease again as a consequence of the new, more recent, mobility restriction measures adopted by Italy. It should be also noticed that, the period following the end of the first stages of the measures limiting physical distance (i.e. after June, the 14$^{th}$) is characterized by a rather slow recovery toward higher values of the mobility. We argue that this result is mainly due to the persistence of a spontaneous attention of people in limiting the number of physical contacts. However, the increasing rate of $D$ becomes suddenly larger in correspondence of October, the 1$^{st}$, likely associated with the combined effect of school opening and resumption of the working activities combined with the reduction of the smart working modality.

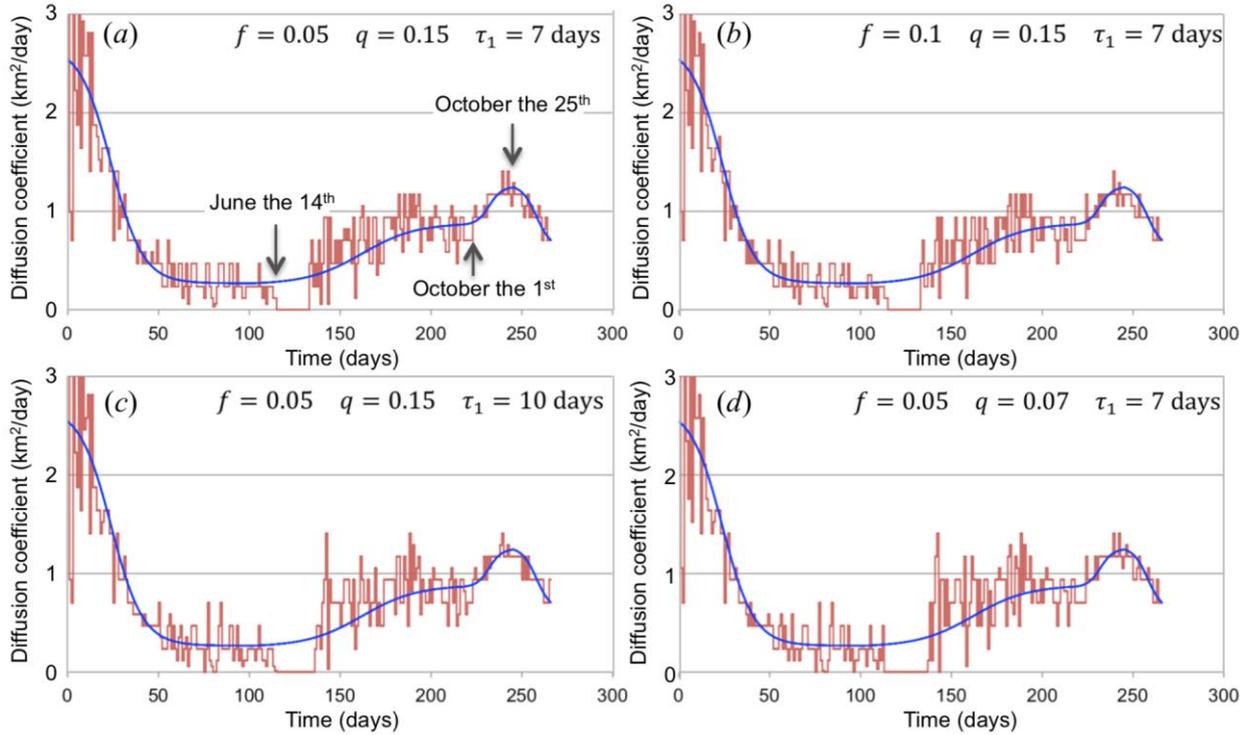

Fig. 2  Diffusion coefficient $D$ as a function of time as determined from the analysis of the data on hospitalized cases shown in Fig. 1, by applying the proposed model (red curve) for different sets of the $f, q, \tau_1$ parameters. The blue continuous curve is a fit to the data by using Eq. 7 and the parameters listed in Tab. I.

The blue line plotted in Fig. 2 is a fit to the $D$ values by using a set of four Fermi–Dirac distribution like functions, each defined within a specific time range, of the kind of:

$$D = D_2 + \frac{D_1 - D_2}{\exp\left(\frac{t - t_0}{\tau_c}\right) + 1} \tag{7}$$

where, for each time range, $D_1$ and $D_2$ are the two levels between which the diffusion coefficient varies in correspondence of any transition associated with the change of the mobility restriction measures, $t_0$ is the time at which the transition occurs, and $\tau_c$ is a measure of the time interval required to change $D$ from the value $D_1$ to $D_2$. We used a unique set of the $D_1, D_2, t_0, \tau_c$ parameter values (reported in Tab. II), to fit all the data shown in Fig. 2, independently of the particular choice of $f$, $q$, and $\tau_1$ [blue continuous line in Fig. 2(a) to 2(d)].

| time range (days) | $0 \leq t < 112$ | $112 \leq t < 212$ | $212 \leq t < 245$ | $t \geq 245$ |
|---|---|---|---|---|
| $D_1$ (km$^2$ days$^{-1}$) | 2.686 | 0.256 | 0.860 | 1.269 |
| $D_2$ (km$^2$ days$^{-1}$) | 0.267 | 0.880 | 1.252 | 0.630 |
| $t_0$ (days) | 24 | 162 | 232 | 258 |
| $\tau_c$ (days) | 8.7 | 15 | 3.5 | 4 |

Tab. I Parameters used to fit the time dependence of the diffusion coefficient given by Eq. 7 to the data shown in Fig. 2.

Once the functional form of $D$ is determined, we can fit the model to the experimental data of Fig. 1, thus getting the best–fit values of $f$, $q$, and $\tau_1$. We focus our fitting procedure to the data concerning the number of both hospitalized case and fatalities (red and grey points in Fig. 1, respectively), since, as we have pointed out, the other information, on positive tested individuals, total infected persons, and healed people, represents only a relatively small fraction of the actual values.

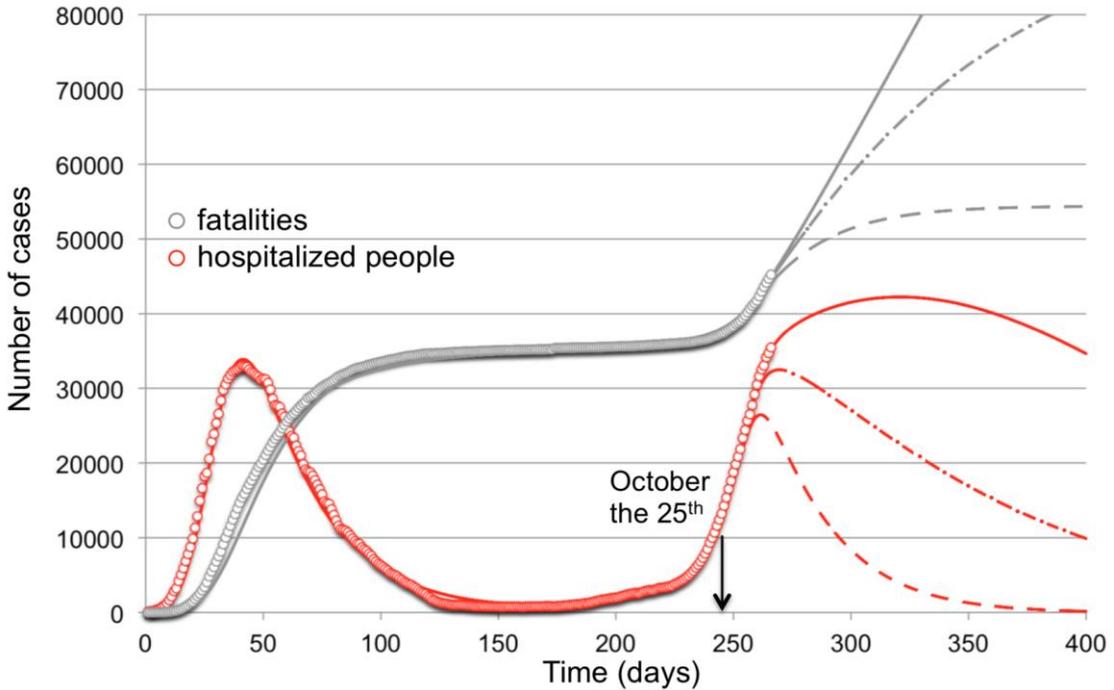

Fig. 3  Linear plot of the reported hospitalized (red circles) and fatalities cases (grey circles). Continuous, dashed, and dot–dashed lines are fits of the model to the data and describe three different scenarios occurring after October the 25$^{th}$, 2020. Dashed lines are the result of calculation under the assumption that the restriction measures adopted on October the 25$^{th}$ could produce a reduction of the diffusion coefficient $D$ to the same level reached as a consequent of the lockdown of March the 9$^{th}$, 2020. Dot–dashed and continuous lines describe the scenarios in which the diffusion coefficient $D$ approaches, after October the 25$^{th}$, 2020, values 2 or 2.36 times larger, respectively.

In addition, in order to improve the fit of the theoretical $r$ and $m$ curves to the corresponding experimental data, we have included the possibility that the fraction $q$ of hospitalized people that unfortunately die (Eq. 3) abruptly moves, after $t = 80$ days, due to the improvement of the effectiveness of the therapeutic protocols, to a value lower than the one characteristic of the first stages of the outbreak diffusion.

| $f$ | $q$ $0 \leq t \leq 80$ | $q$ $t > 80$ | $\tau_1$ (days) |
|---|---|---|---|
| 0.05 | 0.15 | 0.097 | 7.2 |

Tab. II  Parameters used to fit the proposed model (Eqs. 1 to 6) to the experimental data of hospitalized and fatalities cases shown in Fig. 1.

Fig. 3 shows the results of such a fitting procedure (red and grey continuous lines). The corresponding fitting parameters are reported in Tab. II. We notice that the agreement with the experimental data is excellent. Calculations were extended to a time range well beyond the last experimental point ($t = 266$ days) in order to get prediction of the future trend of the number of hospitalized and fatalities cases, accordingly to the more recent mobility restriction measures implemented on October the 25th (the 245th day after the beginning of the outbreak in Italy). Of course, this prediction depends on the particular choice of the parameter values we adopt for describing the time dependence of $D$ for $t \geq 245$ days.

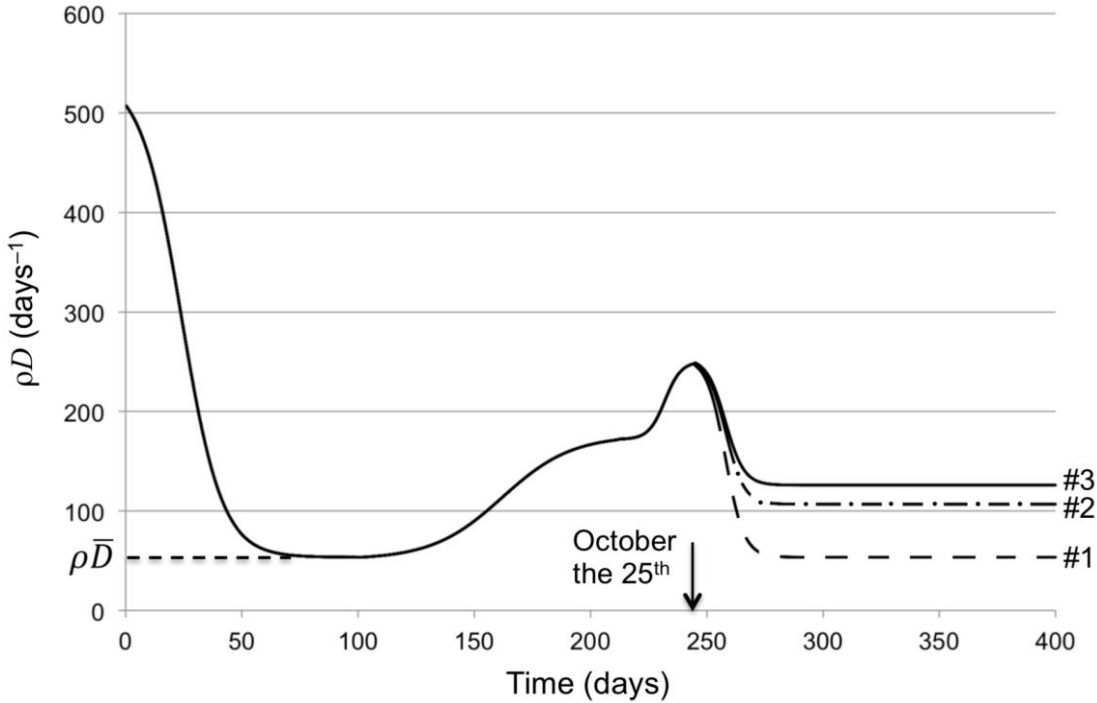

Fig. 4  Time dependence of the function $\rho D$ (average number of people that each individual meets in a day) for the three different scenario simulated as a consequence of the restriction measures adopted on October the 25th, 2020.

In particular, we have put in evidence three different scenarios, simulated by using forms of the $D$ function that differs only after the 245th day (October the 25th, 2020), when the Italian government introduced new measures in order to reduce people mobility. For each of those scenarios, in Fig. 4 the quantity $\rho D$, representing the average number of persons that any individual meets in a day, is plotted as a function of time. In the first scenario, we simulated a situation where $\rho D$ can diminish, for $t \geq 245$ days, quite rapidly ($\tau_c = 4$ days) to the value, $\rho \overline{D}$, that has been reached during the lockdown implemented at the first stages of the outbreak, characterized by particularly severe measures adopted to limit physical distance among people in all the Italian territory. This scenario produces a trend of the hospitalized and fatalities cases that is represented by the dashed lines plotted in Fig. 3, with a number of fatalities that asymptotically approach the value of about $5.4 \times 10^4$. The second scenario is the one according to which $\rho D$ decreases, in the same time range, after the 245$^{th}$ day, to a level two times larger than $\rho \overline{D}$. The consequent behaviour of the hospitalized and fatalities cases is shown as dot–dashed lines in Fig. 3. In that case, the number of fatalities asymptotically approaches the value of about $9.2 \times 10^4$. Last, the third scenario is the one where $\rho D$ reduces, in the same time range, after the 245$^{th}$ day, to a value 2.36 times larger than $\rho \overline{D}$. In this scenario (continuous lines in Fig. 3) the number of fatalities approach to a level close to $1.7 \times 10^5$. The new measures adopted from Italy on October the 25$^{th}$, 2020, introduced light limitation of the people mobility compare to the total lockdown last March. These measures were subsequently strengthened (November the 5$^{th}$, and the 15$^{th}$, 2020), only to a Regional level, depending on the local severity of the outbreak diffusion. To date (November the 15$^{th}$, 2020), only seven Regions out of twenty have reintroduced measures comparable to the full lockdown that has been implemented on March 2020. Although the proposed model does not allow us to follow the spread of infection at a local level, the time evolution of the data on hospitalized and fatalities cases (Fig. 3) is well represented by the third scenario we have discussed, thus indicating the necessity to extend lockdown–like measures to further Regions.

|  | Scenario #1 | Scenario #2 | Scenario #3 |
| --- | --- | --- | --- |
| time range (days) | $t \geq 245$ | $t \geq 245$ | $t \geq 245$ |
| $D_1$ (km$^2$ days$^{-1}$) | 1.269 | 1.269 | 1.269 |
| $D_2$ (km$^2$ days$^{-1}$) | 0.267 | 0.534 | 0.630 |
| $t_0$ (days) | 258 | 258 | 258 |
| $\tau_c$ (days) | 4 | 4 | 4 |

Tab. III Parameters used for determining the time dependence of the diffusion coefficient $D$ according to three different scenarios describing the evolution after the 245$^{th}$ day since the beginning of data reporting on pandemic in Italy.

## Conclusions

In conclusion, we have shown that the spread of COVID–19 can be described by a mean–field diffusion like model. The model successfully reproduces the experimental data on hospitalized and fatalities cases in Italy and allows us to get the time dependence of the diffusion coefficient that influences the dynamics of the outbreak. In particular, the model was successfully used to simulate three different scenarios consequent to the last restriction measures (October the 25th, 2020) of the people mobility in Italy, giving a clear indication of the future trend of the outbreak propagation.